\documentclass[graybox]{svmult}

\usepackage{mathptmx}       
\usepackage{helvet}         
\usepackage{courier}        
\usepackage{type1cm}        
\usepackage{makeidx}         
\usepackage{graphicx}        
\usepackage{multicol}        
\usepackage[bottom]{footmisc}

\usepackage{amssymb,amsmath}
\usepackage[latin1]{inputenc}
\usepackage{cite}
\usepackage{isomath}

\makeindex             

\begin{document}

\title*{Recent advances on information transmission and storage assisted by noise}
\author{P. I. Fierens, G. A. Patterson, A. A. Garc\'ia and D. F. Grosz}
\authorrunning{Fierens et al.} 
\institute{P. I. Fierens \at Instituto Tecnol\'ogico de Buenos Aires (ITBA) and Consejo Nacional de Investigaciones Cient\'ificas y T\'ecnicas (CONICET), Argentina, \email{pfierens@itba.edu.ar}
\and G. A. Patterson \at Facultad de Ciencias Exactas y Naturales -- Universidad de Buenos Aires (FCEN-UBA), Argentina, 
\and A. A. Garc\'ia \at FCEN-UBA, Argentina
\and D. F. Grosz  \at ITBA and CONICET, Argentina
}
%
%
\maketitle

\abstract{The interplay between nonlinear dynamic systems and noise has proved to be of great relevance in several application areas. In this presentation, we focus on the areas of information transmission and storage. We review some recent results on information transmission through nonlinear channels assisted by noise. We also present recent proposals of memory devices in which noise plays an essential role. Finally, we discuss new results on the influence of noise in memristors.}

\section{Introduction}
\label{sec:intro}

In communication systems noise is usually regarded as a nuisance to cope with. However, it has been shown that in certain nonlinear channels information transmission is actually sustained by noise (see, e.g., \cite{Chapeau-Blondeau.1999,Chapeau-Blondeau.2000,Lindner.PhysRevLett.1998,Locher.PhysRevLett.1998,Garcia-Ojalvo.EPL.2000}). In particular, dynamic systems comprised of a chain of bistable double-well potentials driven by a periodic signal have been extensively studied and were shown to be able to sustain noise-assisted fault-tolerant transmission \cite{Perazzo.PhysRevE.2000, Zhang.PhysRevE.1998}. In Section \ref{sec:transmission} we review some recent progress on this topic.

In the last few years there has been an increased effort in the search of alternative technologies for computer memory devices. This effort is motivated by the perceived near future end of the ability of current technologies to provide support for the exponential increase of memory capacities as predicted by Moore's law. In this context, logic gates that work with the help of noise have been suggested (see, e.g., \cite{Murali.PRL.2009,Murali.APL.2009,Murali.ICAND.2010,Bulsara.ChemicalPhysics.2010,Guerra.NanoLett.2010,
Singh.PhysRevE.2011,Dari.BioCAS.2011,Dari.PhysRevE.2011,Dari.EPL.2011}). In a similar vein, it is possible to think of memories which can benefit from, and indeed work only in, the presence of noise. In Section \ref{sec:storage}, we review some recent advances in this area.

Resistive memories represent one of the most promising candidates for next generation computer memories, and are usually associated with a type of two-terminal passive circuit element known as a {\it memristor} \cite{Strukov.Nature.2008,Chua.IEEEProc.2012}. Recently, Stotland and Di Ventra \cite{Stotland.PRE.2012} showed that noise may help increase the contrast ratio between low and high memory states. In Section \ref{sec:memristor}, we present some new experimental results on the role played by noise in this type of system, extending those in Ref. \cite{Stotland.PRE.2012}.

\section{Information transmission}
\label{sec:transmission}

System performance in transmission lines can be characterized by means of several metrics commonly encountered in the area of communications, e.g., output Bit Error Rate (BER) and Signal-to-Noise Ratio (SNR) (see, e.g., \cite{Sklar,Agrawal}). The BER of a communication system is a measure of the probability of receiving erroneous bits and represents one of the most important performance metrics used in digital communications. The SNR is also an important metric in analogue communications and its relevance in digital systems stems from the usual monotonic relation between BER and SNR. Specifically, for an additive Gaussian noise (AWGN) channel, increasing the SNR decreases the minimum allowable bit error rate \cite{Blahut.1987}. However, this conclusion is not necessarily valid for a nonlinear communication channel. Several authors have studied bit error rate metrics for the case of a single double-well potential such as the one described in 
\cite{McNamara.PhysRevA.1989,Zhang.PhysRevE.1998,Perazzo.PhysRevE.2000}. Barbay et al.\cite{Barbay.PhysRevE.2001,Barbay.PhysRevLett.2000} developed theoretical expressions for the BER performance of a VCSEL which were experimentally validated. Duan et al. \cite{Duan.PhysRevE.2004} also presented a theoretical calculation of the BER for a supra-threshold signal strength. Moreover, Godivier and Chapeau-Blondeau \cite{Godivier.ELetters.1997} studied the capacity of a channel comprised of a single double-well potential.

Ibáñez et al. \cite{Ibanez.PhysD.2009} studied a double-well forward-coupled information transmission line. In this type of nonlinear channel, two different transmission regimes were identified, namely noise-supported and coupling-supported, corresponding to sub-critical and super-critical coupling strengths, respectively. While addition of noise to each potential well is required in order to sustain transmission in the sub-critical case, a super-critical coupling is strong enough to guarantee operation of the transmission line even in the absence of noise. In Ref. \cite{Ibanez.PhysD.2009}, it was shown that output BERs remain flat for a broad range of added noise, especially for a super-critical coupling strength. 

\begin{figure}[hb]
\begin{center}
\includegraphics[scale=0.85]{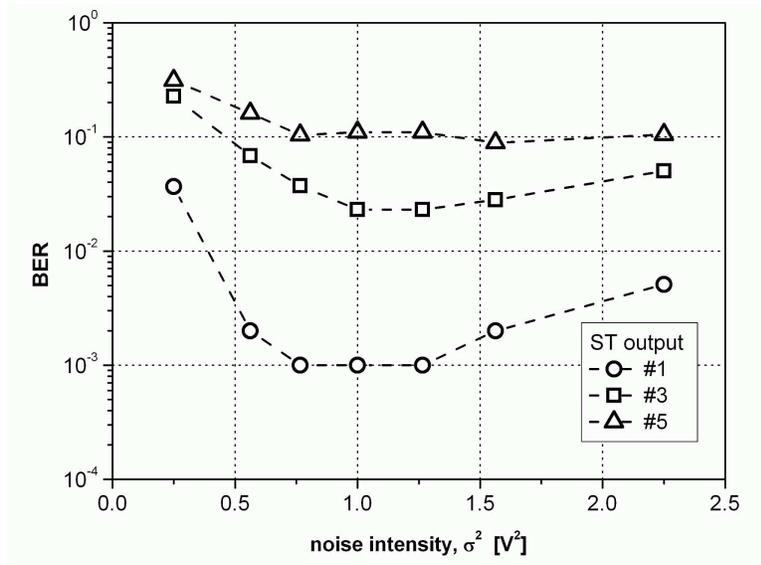}
\end{center}
\caption{Bit error rate as a function of noise intensity in a chain of five Schmitt triggers. BER is minimized for an optimal noise intensity. Figure taken from Ref. \cite{Patterson.PhysA.2010}.}
\label{fig:ber_vs_noise}
\end{figure}

\begin{figure}[ht]
\begin{center}
\includegraphics[scale=0.85]{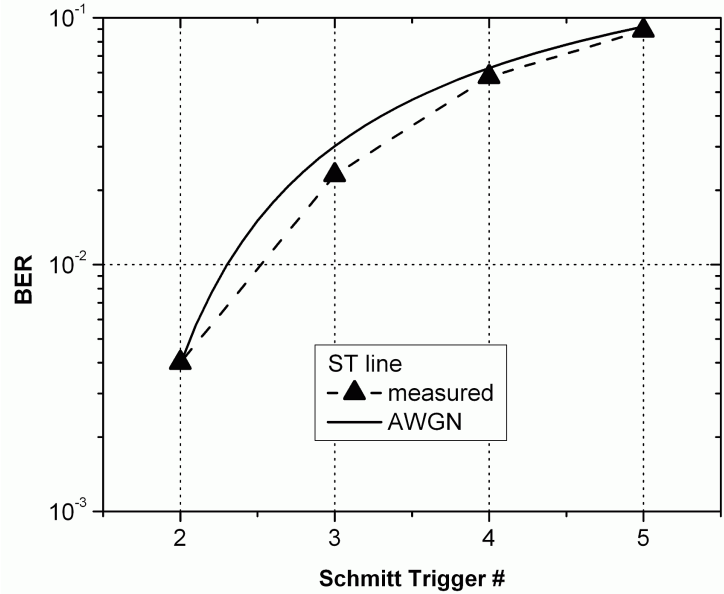}
\end{center}
\caption{Performance of a transmission line comprised of five in-line Schmitt triggers. The performance expected from an additive white Gaussian noise channel is also shown as a reference. Figure taken from Ref. \cite{Patterson.PhysA.2010}.}
\label{fig:ber_awgn}
\end{figure}

Schmitt triggers (STs) are commonly used as `discrete' models of double-well potentials (see, e.g, \cite{McNamara.PhysRevA.1989}). They also can be used as simple models of lasers with saturable absorbers, which are known to exhibit a stochastic resonant behavior \cite{Gammaitoni.RevModPhys.1998}. As some schemes for optical pulse amplification and shaping (known as ``2R regeneration'') mimic fast saturable-absorbers \cite{Mamyshev.1998}, STs may serve as experimental toy models for the analysis of some forms of optical regeneration in communication systems. A similar approach was followed, e.g., by Korman et al. \cite{Korman.Siam.2002} who modeled   a mid-span repeater with a hysteresis-type nonlinearity of the same kind found in a Schmitt trigger. Patterson et al. \cite{Patterson.PhysA.2010} performed an experimental investigation of the transmission properties of a line comprised of five in-series Schmitt triggers from the point of view of a communication system, where each ST was fed with white Gaussian noise, and the first ST was driven by a pseudo-random sequence of bits. As shown in Fig. \ref{fig:ber_vs_noise}, an optimal noise intensity was found for which the BER is minimized. The rate of system performance degradation with distance (number of STs traversed by the information-carrying signal) was found to be similar to that observed in a linear AWGN communication channel (see Fig. \ref{fig:ber_awgn}). 

A transmission line has an associated delay which depends on its physical characteristics. The time-delay properties of a double-well forward-coupled information transmission line were investigated in Ref. \cite{Ibanez.FNL.2008}, showing that it can be regarded as a noise-tunable delay line for a broad range of noise and coupling-strength parameters. Such a tunable delay line may find applications, e.g., in the phase modulation of information-carrying signals. 

\section{Information storage}
\label{sec:storage}

\begin{figure}[ht]
\begin{center}
\includegraphics[scale=0.35]{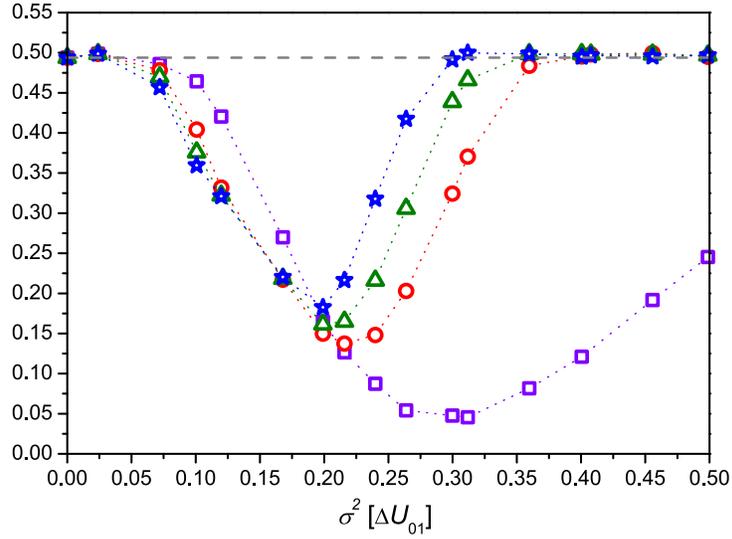}
\end{center}
\label{fig:memory_two_error}
\caption{Probability of erroneous information retrieval from the second oscillator as function of noise intensity for different observation times ($T=T_P$ - squares, $10 T_P$ - circles, $20T_P$ - triangles, $40T_P$ - stars). $T_P$ is an arbitrary time scale. Figure taken from Ref. \cite{Ibanez.EPJB.2010}.}
\end{figure}

\begin{figure}[hb]
\begin{center}
\includegraphics[scale=0.35]{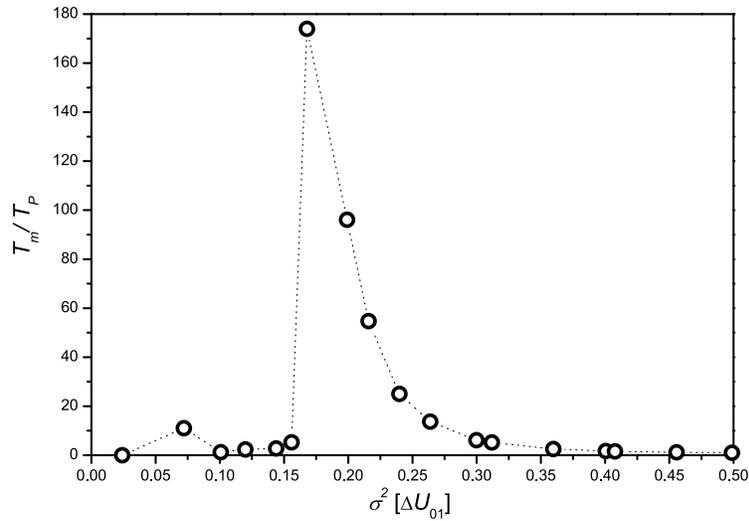}
\end{center}
\label{fig:memory_persistence}
\caption{Memory persistence as a function noise intensity. There is a noise intensity that maximizes persistence. Figure taken from Ref. \cite{Ibanez.EPJB.2010}.}
\end{figure}

Carusela et al. \cite{Carusela.PhysRevE.2001,Carusela.PhysicaD.2002} showed that a double-well forward-coupled transmission line, such as that described in Section \ref{sec:transmission}, can work as a memory element when closed onto itself forming a loop. In particular, they showed that such a ring was able to sustain a traveling wave with the aid of noise long after the harmonic drive signal had been switched off. Building on the work in Refs. \cite{Carusela.PhysRevE.2001,Carusela.PhysicaD.2002}, Ibáñez et al. \cite{Ibanez.EPJB.2010,Fierens.PhysLettA.2010} showed that a ring of two bistable oscillators is capable of storing a single bit of information in the presence of noise. Fig. \ref{fig:memory_two_error} shows the probability of erroneous retrieval from the second oscillator. It can be observed that, for each retrieval time, there is an optimal noise intensity for which the probability of error is minimized. In Ref. \cite{Ibanez.EPJB.2010} memory persistence time ($T_m$) was defined as the time elapsed until the first oscillator reaches a probability of error equal to that of the noiseless case. Fig. \ref{fig:memory_persistence} shows that there is an optimal noise intensity which maximizes memory persistence. Foreseeing a practical implementation of this type of memory device, its performance under varying physical parameters, such as noise bandwidth and time response of the bistable elements, was investigated in Ref. \cite{Fierens.ICAND.2010}. In particular, it was found that the device is more resilient to the action of noise when the noise and the bistable element bandwidths are of the same magnitude.

Bellomo et al. \cite{Bellomo.PhysLettA.2011} proposed a multibit storage device consisting on a single Schmitt trigger (ST) and an element that introduces a finite delay in a loop configuration. The proposed device can be considered as a toy model of a long transmission link with in-line nonlinear elements such as saturable absorbers (\cite{Mamyshev.1998,Gammaitoni.RevModPhys.1998}). Nonlinear delayed loops have been extensively studied (see, e.g., \cite{Aida.IEEEQuantumE.1992,Losson.Chaos.1993,Mensour.PhysLettA.1995,Mensour.PhysLettA.1998,Morse.PhysLettA.2006,Jin.PhysicaA.2007,Middleton.PhysRevE.2007}). This type of systems usually presents a complex nonlinear behavior, including self-sustained oscillations and chaotic operation regimes. Memory devices that make use of regimes that show multistable behavior of delayed feedback loops have been proposed. Ref. \cite{Aida.IEEEQuantumE.1992} presented a memory device that stores bits coded as particular oscillation modes of a delay feedback loop with an electro-optical modulator. Similarly, in Ref. \cite{Mensour.PhysLettA.1995} was shown that binary messages can be stored using controlled unstable periodic orbits of a particular class of delay-loop differential equations. However, memory devices proposed in Refs. \cite{Aida.IEEEQuantumE.1992, Mensour.PhysLettA.1995} were not assisted by noise. Refs. \cite{Mensour.PhysLettA.1998} and \cite{Jin.PhysicaA.2007} studied the behavior of a delayed loop with a single threshold device and a bistable device, respectively, but focusing on the response of the system to a harmonic driving signal.

\begin{figure}[hb]
\begin{center}
\includegraphics[scale=0.35]{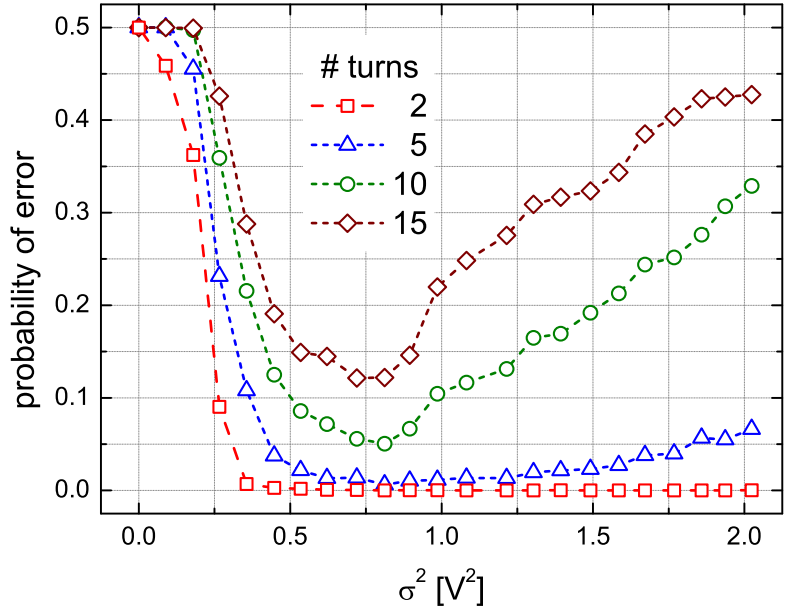}
\end{center}
\label{fig:memory_multibit}
\caption{Probability of erroneous retrieval as a function of noise intensity when 4 bits are stored. Different curves correspond to varying elapsed times. A turn corresponds to the mean time it takes the signal to travel through the loop elements. Figure taken from Ref. \cite{Bellomo.PhysLettA.2011}.}
\end{figure}

\begin{figure}[ht]
\begin{center}
\includegraphics[scale=0.3]{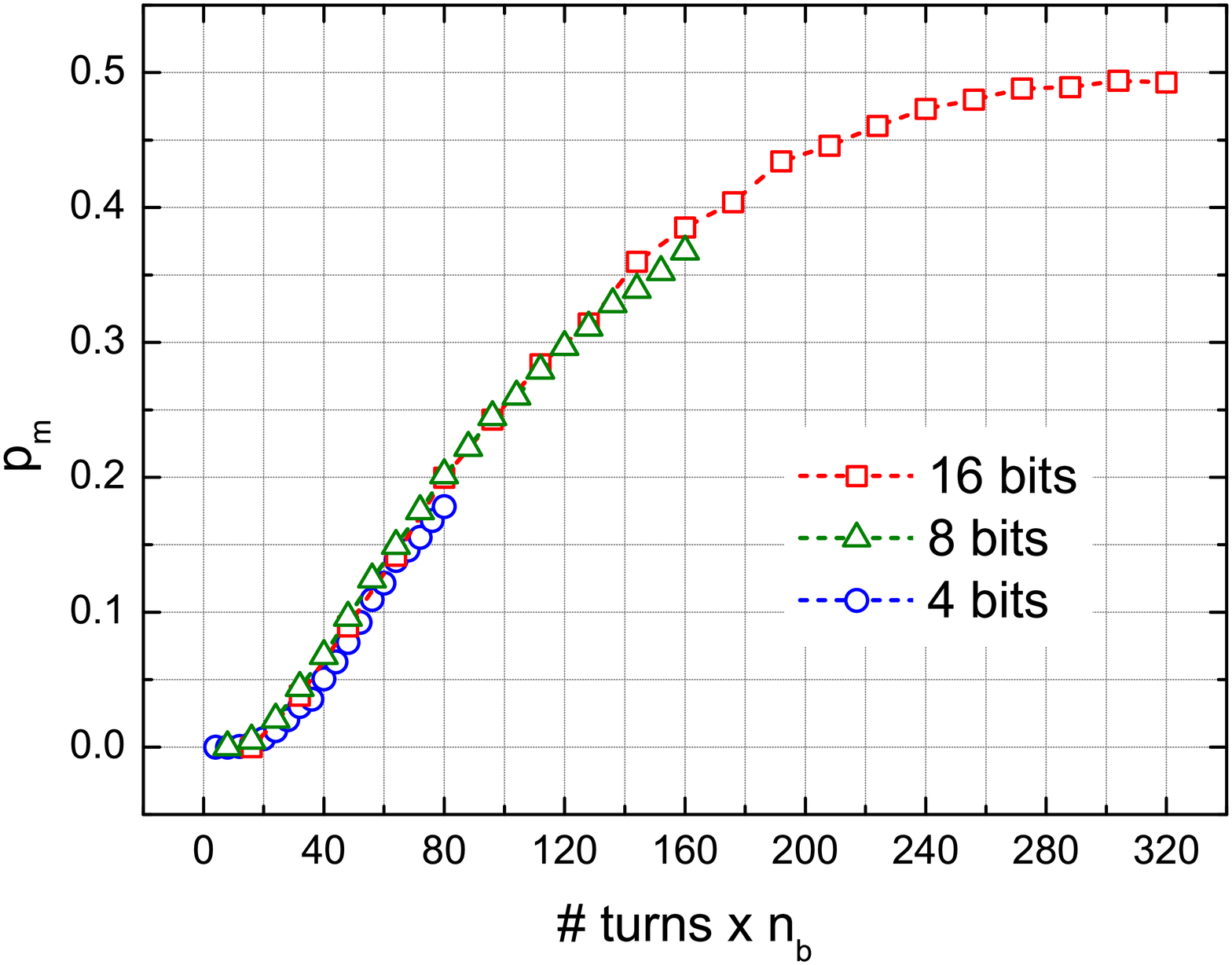}
\end{center}
\label{fig:memory_multibit_normalized}
\caption{Minimum probability of error as a function of elapsed time normalized to the bit duration. $n_b$: number of stored bits. Figure taken from Ref. \cite{Bellomo.PhysLettA.2011}.}
\end{figure}

Experimental studies in Ref. \cite{Bellomo.PhysLettA.2011} showed that the performance of the proposed multibit memory device is optimal for an intermediate value of noise intensity (see Fig. \ref{fig:memory_multibit}). It is interesting to note that, although performance was shown to deteriorate with time and with the number of stored bits, it also was found that the probability of error is independent of the number of bits when the elapsed time is normalized to the bit duration (see Fig. \ref{fig:memory_multibit_normalized}), a fact of relevance when considering practical implementations of the device. 	

\section{Noise in memristors}
\label{sec:memristor}

One of the proposed alternatives for succeeding current nonvolatile memory technologies is the so-called family of Resistive Random-Access Memories (ReRAMs) (see, e.g., \cite{Sawa.MaterialsToday.2008}). ReRAMs are based in the resistive switching phenomenon observed in several materials, that is, the change of electrical resistance by the application of electrical pulses. Indeed, a high resistance may represent, say, a `0', and a low resistance may represent a `1' logic state.

Often, resistive switching devices are associated with a type of two-terminal passive circuit element known as a {\it memristor} \cite{Strukov.Nature.2008,Chua.IEEEProc.2012} which was originally proposed by Chua in Ref. \cite{Chua.IEEETransCircuitTheory.1971}. There are different proposed mechanisms which try to explain the observed behavior of resistive switching materials (see, e.g.,
\cite{Sawa.MaterialsToday.2008,Strukov.Nature.2008,Pickett.JournalofAppliedPhysics.2009,Sanchez.APL.2007,Rozenberg.PhysRevB.2010,Ghenzi.JournalAppliedPhysics.2010,Gomez-Marlasca.APL.2011} and references therein). From a macroscopic point of view, one of the simplest models is that proposed in Ref. \cite{Strukov.Nature.2008}, described by
\begin{equation}
v(t) = R(s)i(t),
\label{eq:ohm}
\end{equation}
\begin{equation}
\frac{ds}{dt} = \alpha F(s,i)i(t),
\label{eq:memristor_state}
\end{equation}
where $v(t)$ is the applied voltage, $i(t)$ is the current, $R$ is the device's resistance, $s\in[0,1]$ is an internal state variable, $\alpha$ is a constant and $F(s,i)$ is a nonlinear function. Among the several suggested alternatives for $F(s,i)$ (see, e.g., \cite{Joglekar.EuropeanJournalofPhysics.2009}), one commonly used is
\begin{equation}
F(s,i) = 4s(1-s).
\label{eq:memristor_window}
\end{equation}
Based on these equations, Stotland and Di Ventra \cite{Stotland.PRE.2012} showed that noise may help increase the contrast ratio between low and high resistance values. This conclusion has important practical consequences as the contrast in resistance values can be associated with the probability of error in ReRAMs. Stotland and Di Ventra modified Eq. \ref{eq:memristor_state} by adding an internal noise term,
\begin{equation}
\frac{ds}{dt} = \alpha F(s,i)i(t)+\eta(t),
\label{eq:memristor_state_ext_noise}
\end{equation}
where $\eta(t)$ is white Gaussian noise. Patterson et al. \cite{Patterson.ICAND.2012} extend the results of Stotland and Di Ventra \cite{Stotland.PRE.2012}. In particular, they show that external noise, that is, noise added to the externally applied voltage, does not help increase the contrast between low and high resistance states if Eqs. \ref{eq:ohm}-\ref{eq:memristor_window} are valid.

In order to explore the behavior of memristors under externally applied noise, we experimentally studied a sample of a manganite, $\mathrm{La}_{0.35}\mathrm{Pr}_{0.300}\mathrm{Ca}_{0.375}\mathrm{Mn}\mathrm{O}_3$. The behavior of this material has already been studied in, e.g., \cite{GomezMarlasca.JournalPhysics.2009,Ghenzi.JournalAppliedPhysics.2010,Gomez-Marlasca.APL.2011,Ghenzi.JournalAppliedPhysics.2012}. Two silver contacts were placed on the sample and an externally controlled current was applied through them. The resistance between contacts was calculated by measuring the voltage drop across them when a small ($\pm 1$ mA) noiseless current was applied. Current pulses of $1$ ms of duration were applied with varying amplitudes. Each current pulse was followed by 1 second of `silent' (no current applied) `setting' time and preceded by $20$ ms of the small current used for resistance measurement ($-1$ mA applied during $10$ ms and $+1$ mA during other $10$ ms). Samples of zero mean white Gaussian current noise with $100$ mA standard deviation were generated by software and added to the externally applied current. The actual noise bandwidth and detailed characteristics depend on several elements of the experimental setup such as, e.g., the bandwidth and the linearity of the digital-to-analog converter.

\begin{figure}[h]
\begin{center}
\includegraphics[scale=0.35]{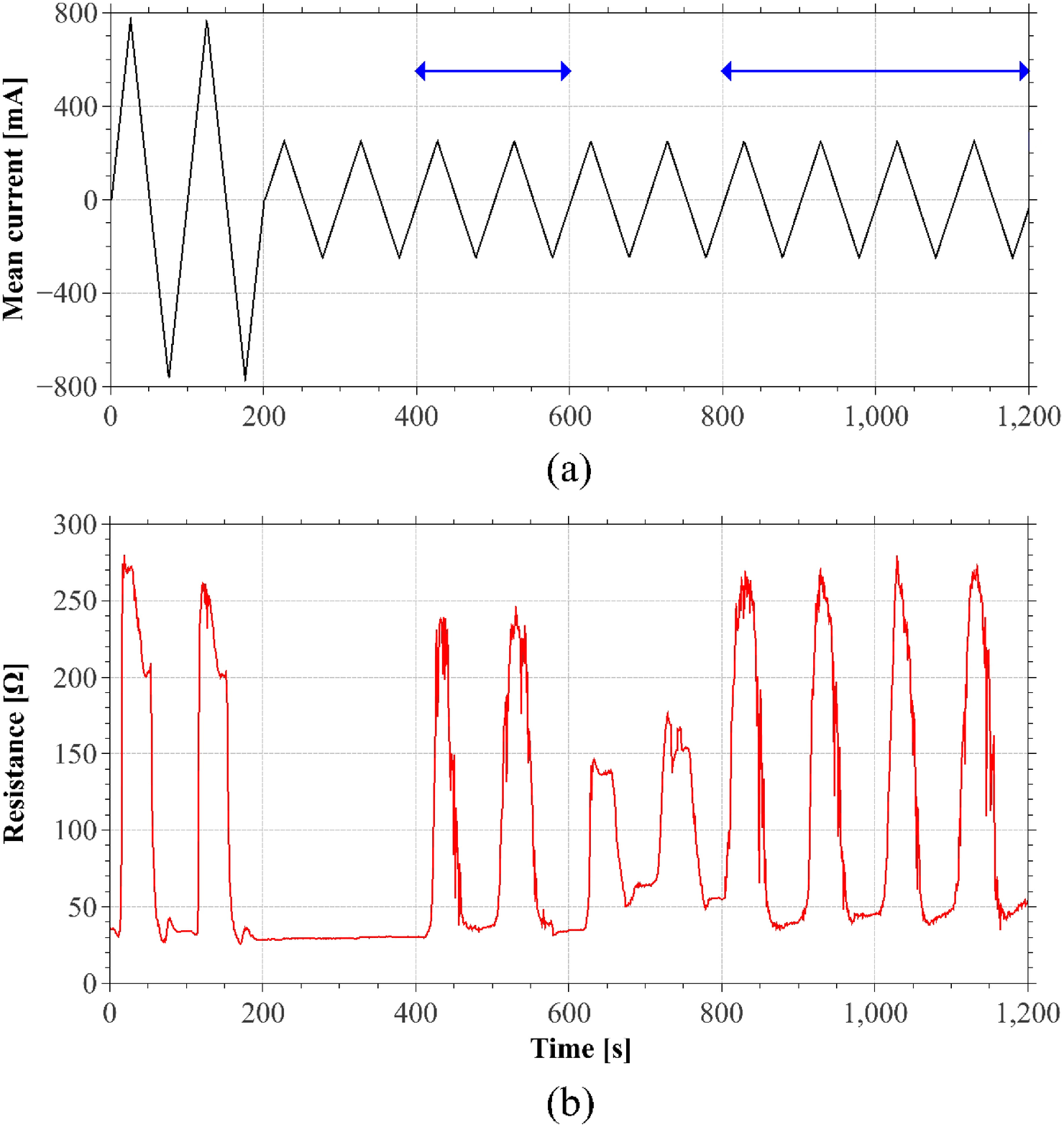}
\end{center}
\label{fig:exp_data}
\caption{Experimental results on the effect of noise on a memristor. (a): measured mean pulse current as a function of time. (b): measured memristor's resistance. Intervals where noise was added are indicated by (blue) arrows in (a).}
\end{figure}

Fig. \ref{fig:exp_data} shows the results of a typical experiment. For the first $200$ seconds, noiseless high-current pulses (up to $800$ mA) were applied and the memristor's resistance varied between $\sim 30\mathrm{\Omega}$ and $\sim 270\mathrm{\Omega}$. During the following $200$ seconds, the current amplitude was lowered (down to $\sim 200$ mA) and no noticeable change on the resistance was observed. The low amplitude pulses were repeated for other $200$ seconds, but now noise was added. Interestingly, the memristor responded to the added noise by varying its resistance in almost the same magnitude as it did when high-current pulses were applied. When noise is turned off, the memristor's resistance varies, but the change is not of the same magnitude as when noise is applied.

In summary, experimental data confirms that externally applied noise increases the contrast between low and high resistance states in a memristor, a fact of relevance in applications. Considering that Eqs. \ref{eq:ohm}-\ref{eq:memristor_window} disallow such an effect of noise (see \cite{Patterson.ICAND.2012}), our experimental results also point toward the need for more complex models.

\section{Conclusions and future work}

We have reviewed some recent advances on the field of information transmission and storage assisted by noise. Schemes of in-series hysteretic nonlinear elements, modeled by Schmitt triggers, were shown to enable transmission of subthreshold signals with low probability of errors for optimal noise intensities. Closed loops of such elements were also shown to sustain information storage only in the presence of moderate amounts of noise. Future work in these areas should address the influence of different types of noise. For example, 1/f-noise is found in several applications (see, e.g., \cite{Dutta.RMP.1981}) and its eventual beneficial role in nonlinear systems as those described in this work has not been deeply investigated yet. It also remains to be explored the possibility of practical applications of noise-tunable delay lines as those described in Section \ref{sec:transmission}.

We have also presented results on the effect of noise in memristors, where external noise helps commute resistive states in the presence of a small amplitude driving field. We believe these results are of significance since memristors are expected to operate in electronic circuits with a large scale of integration, and as such the effects of thermal noise will have to be coped with. Since simple models (see \cite{Patterson.ICAND.2012}) cannot account for the presented results, future work should focus on more complete models of memristors behavior.

\begin{acknowledgement}
We gratefully acknowledge financial support from ANPCyT under project PICT-2010 \#121. Memristor samples were provided by the Laboratory of Electrical and Magnetic Properties, Condensed Matter Group, CNEA (Argentina). We want to specially thank Fernando Gomez-Marlasca and Pablo Levy for their suggestions on the experiments.
\end{acknowledgement}

\bibliographystyle{spphys}
\bibliography{biblio}

\end{document}